\documentclass[aps,twocolumn,showpacs,showkeys,floatfix]{revtex4}
\usepackage{graphicx}% Include figure files
\usepackage{dcolumn}% Align table columns on decimal point
\usepackage{bm}% bold math

\newcommand{\dfrac}[2]{\displaystyle\frac{#1}{#2}}

\begin{document}

%\preprint{Manuscript ??????}

\title{On High Explosive Launching of Projectiles for Shock Physics Experiments}

\date{February 27, 2007 -- LA-UR-07-1299}

\author{Damian C. Swift}
\email{dswift@lanl.gov}
%\homepage{http://public.lanl.gov/dswift}
\affiliation{%
   Group P-24, Physics Division, Los Alamos National Laboratory,
   MS~E526, Los Alamos, New Mexico 87545, USA
}

\author{Charles A. Forest}
%\email{caforest@losalamos.com}
\affiliation{%
%  2247, 34th Street, - if OK'd
   Los Alamos, New Mexico 87544, USA
}

\author{David A. Clark}
%\email{daclark@lanl.gov}
\affiliation{%
   Group P-22, Physics Division, Los Alamos National Laboratory,
   MS~D410, Los Alamos, New Mexico 87545, USA
}

\author{William T. Buttler}
% Craig McCluskey??
%\email{buttler@lanl.gov}
\affiliation{%
   Group P-23, Physics Division, Los Alamos National Laboratory,
   MS~H803, Los Alamos, New Mexico 87545, USA
}

\author{Mark Marr-Lyon}
%\email{mmarr@lanl.gov}
\affiliation{%
   Group HX-3, Hydrodynamic Experimentation Division, Los Alamos National Laboratory,
   MS~P940, Los Alamos, New Mexico 87545, USA
}

\author{Paul Rightley}
%\email{prightley@lanl.gov}
\affiliation{%
   Group HX-3, Hydrodynamic Experimentation Division, Los Alamos National Laboratory,
   MS~P940, Los Alamos, New Mexico 87545, USA
}

\begin{abstract}
The hydrodynamic operation of the `Forest Flyer' type of explosive
launching system for shock physics projectiles was investigated in detail using
one- and two-dimensional continuum dynamics simulations.
The simulations were numerically converged and insensitive to uncertainties
in the material properties; they reproduced the speed of the projectile and
the shape of its rear surface.
The most commonly-used variant, with an Al alloy case, 
was predicted to produce a slightly curved projectile,
subjected to some shock heating, and
likely exhibiting some porosity from tensile damage.
The curvature is caused by a shock reflected from the case;
tensile damage is caused by the interaction of the Taylor wave pressure 
profile from the detonation wave with the free surface of the projectile.
The simulations gave only an indication of
tensile damage in the projectile, as damage is not understood well enough
for predictions in this loading regime.
The flatness can be improved by using a case of lower shock impedance,
such as polymethyl methacrylate.
High-impedance cases, including Al alloys but with denser materials 
improving the launching efficiency, can be used if designed according to
the physics of oblique shock reflection, which indicates an appropriate
case taper for any combination of explosive and case material.
The tensile stress induced in the projectile depends on the relative
thickness of the explosive, expansion gap, and projectile.
The thinner the projectile with respect to the explosive, the smaller the
tensile stress.
Thus if the explosive is initiated with a plane wave lens, the tensile
stress is lower than for initiation with multiple detonators over a plane.
The previous plane wave lens designs did however induce a tensile stress
close to the spall strength of the projectile.
The tensile stress can be reduced by changes in the component thicknesses.
Experiments to verify the operation of explosively-launched projectiles
should attempt to measure porosity induced in the projectile:
arrival time measurements are likely to be insensitive to porous regions
caused by damaged or recollected material.
\end{abstract}

% 06.60.Jn High-speed techniques (microsecond to femtosecond)
% 07.35.+k High-pressure apparatus; shock tubes; diamond anvil cells
%% 46.70.-p Application of continuum mechanics to structures
%% 46.70.De Beams, plates and shells
% 47.40.-x Compressible flows; shock and detonation phenomena
% 62.50.+p High-pressure and shock wave effects in solids and liquids
%\keywords{Suggested keywords}%Use showkeys class option for keyword display

\maketitle

\section{Introduction}
The impact of a disk-shaped projectile against a flat target 
induces shock waves in the target and the projectile, by which high pressures 
can be generated and investigated \cite{Bushman93}.
The projectile may be accelerated in various ways, including
by the expansion of compressed gas or hot products of a chemical reaction,
electromagnetic forces from a pulsed electrical discharge,
or the expansion of material heated by a laser.
With experiments involving the impact of a projectile with a target, 
a frequently-encountered problem is the synchronization of measurements
with the short-lived shocked state.
The duration of the shocked state is very short compared with the time
required to accelerate the projectile, so any uncertainty in
triggering time, projectile speed, or acceleration distance makes 
measurements disproportionately difficult.

Chemical reaction products are most commonly used in a powder gun,
where the chemical species react by deflagration, and acceleration
takes place over a few meters as the projectile is accelerated along a barrel.
The acceleration distance can be reduced greatly if the chemical species
react by detonation instead
\cite{Bushman93,Forest98}, i.e. by the use of explosives to accelerate the
projectile.
Explosives can also make it easier to reach higher speeds and therefore
higher impact pressures.

Explosively-launched projectiles have been used to induce shock states with
relatively easy synchronization with the novel technique of neutron
resonance spectroscopy (NRS),
where a pulse of neutrons generated from a spallation target was
used to measure the temperature of the target material while it was
in the shocked state \cite{Yuan05}.
These experiments gave unexpected temperatures when performed on Mo
as a reference material,
which can be explained largely by the sensitivity of the NRS technique
to subtleties of the loading conditions induced by the projectile
\cite{Swift_nrs_07}.
The purpose of this paper is to review the physics of a class of 
explosively-launched
projectiles in more detail, identifying aspects of the design which may
cause the impact-induced loading to deviate from a one-dimensional shock,
and demonstrating palliative measures for improved designs.

\section{`Forest Flyer' design}
\begin{figure*}
\begin{center}
\includegraphics[scale=0.3]{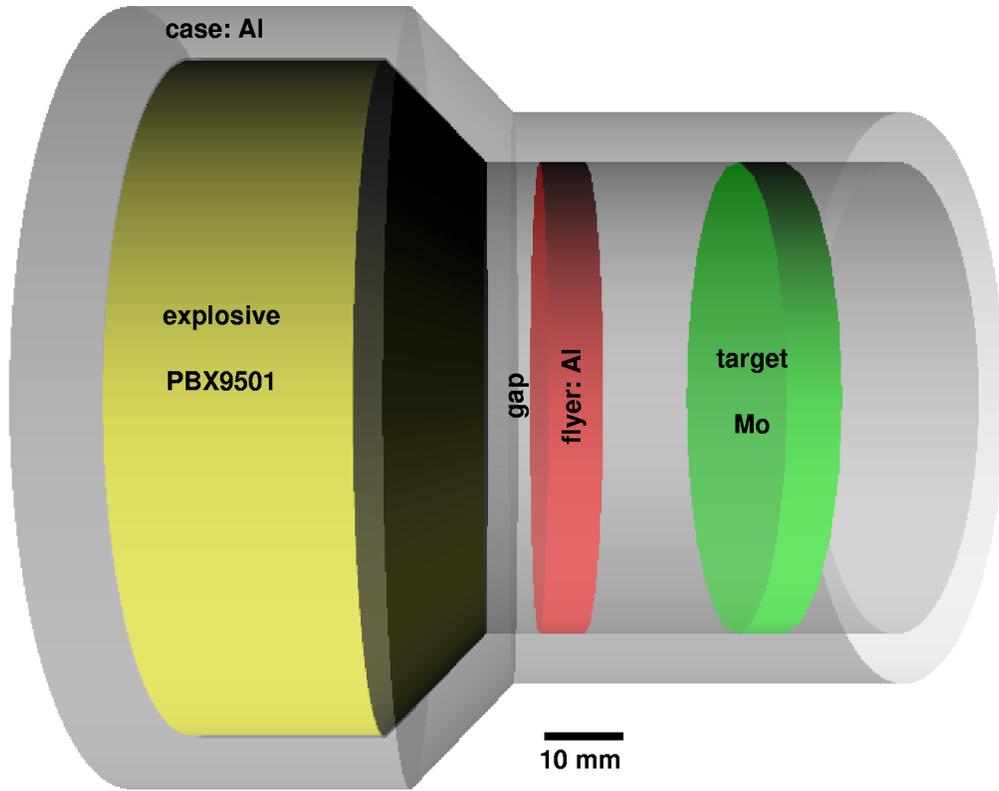}
\end{center}
\caption{Schematic of `Forest Flyer' explosive launching system for
   shock physics projectiles.}
\label{fig:forest_schem}
\end{figure*}

The explosively-launched projectiles used in the NRS experiments followed
the `Forest Flyer' design \cite{Forest98}.
In this design, the explosive charge was initiated across a planar face
and the detonation products expanded across a gap before accelerating the
projectile.
If the gap were omitted, the detonation wave would drive a strong shock
into the projectile, inducing shock heating and strong tensile stresses
after interacting with the free surface, which could cause damage and
spallation.
The explosive charge was tapered down to the diameter of the projectile.
(Fig.~\ref{fig:forest_schem}.)

\begin{figure*}
\begin{center}
\includegraphics[scale=0.5]{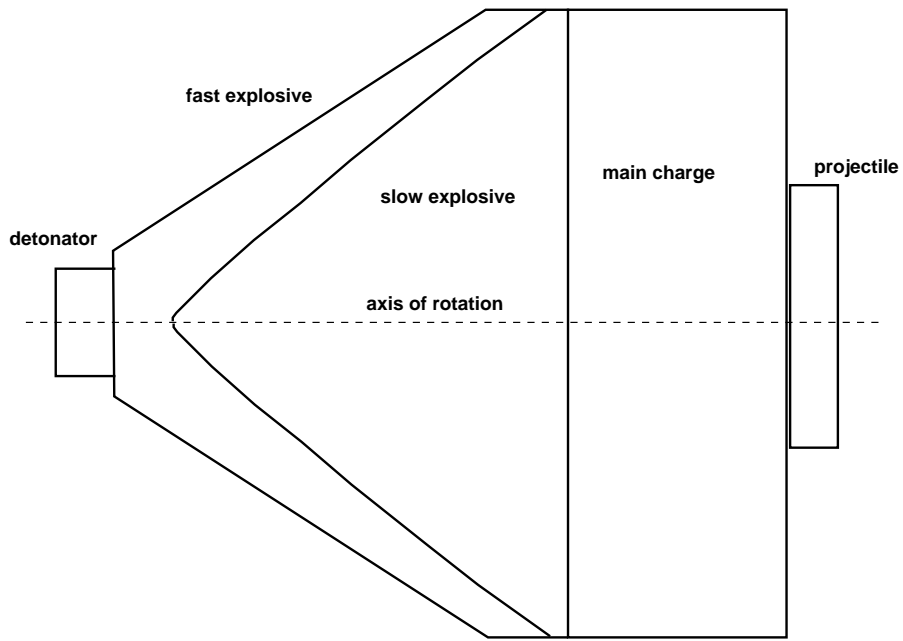}
\end{center}
\caption{Schematic of boosted plane wave lens launching system for
   shock physics projectiles.}
\label{fig:pwlflyer_schem}
\end{figure*}

The Forest Flyer design was originally developed to improve on older
designs which used a cylindrical explosive charge, initiated by a 
plane wave lens, to launch a disk-shaped projectile 
(Fig.~\ref{fig:pwlflyer_schem}).
The projectile had a smaller diameter than the charge, so a significant
part of the explosive energy did not contribute to the acceleration of the
projectile.
This is particularly inefficient if it is desired to perform a shock wave
experiment inside a containment vessel -- for example, for experiments on
hazardous materials -- as the mass of explosive should be minimized.
Another disadvantage is that reaction products of the explosive blow past
the projectile, and may interfere with the experiment or diagnostics.
The Forest Flyer design improved over the older design by removing
explosive which did not contribute to the acceleration of the projectile,
and adding a case which confined the remaining explosive,
increasing the efficiency of the projectile launch and protecting the
impact experiment from interference by reaction products.

The Forest Flyer design was documented as a conference proceedings paper
\cite{Forest98} of a poster presentation.
Unfortunately, this reference contains some errors which are corrected
and clarified here.
The explosive charge and case were tapered down to the diameter of the 
projectile at an angle intended to avoid the case affecting the shape
of the detonation wave in the charge \cite{Aveille89}.
If the case is made of a material of higher shock impedance than the
explosive, it will reflect a shock wave which may make the detonation
run faster close to the case; if the material has a lower shock impedance,
it will reflect a rarefaction which may make the detonation wave run slower.
The taper angle can be chosen to minimize the perturbation of a given
case material on the detonation wave.
The previously-published analysis \cite{Forest98} applies to cases of
lower impedance, though variants of the Forest Flyer have been used with
cases of higher impedance, such as Al alloys.
In fact, the Forest Flyer has been used generally with the high
explosive PBX-9501, comprising mainly cyclotetramethylene tetranitramine (HMX), 
whose detonation wave is relatively insensitive to
the case \cite{Collyer98}.
A more serious concern is the potential for deformation of the projectile
caused by any shock reflected into the detonation products by the taper,
for a high-impedance case.

Variants of the Forest Flyer design have been used for a range of shock
physics experiments.
Forest Flyer projectiles have been used to induce shock waves in Mo and
overdriven detonation waves in PBX-9501, to test the use of neutron resonance
spectrometry as an internal temperature diagnostic for opaque materials
\cite{Yuan05}.
These experiments were performed in a restrictive containment vessel, 
so a multipoint detonator was used in preference to a plane wave lens 
to initiate the explosive charge, minimizing the mass of explosive.
Forest Flyer projectiles have been used to induce surface ejecta from Sn
samples, to study the effect of surface finish
\cite{Buttler06}.
In these experiments, the projectile and case were Al alloy.

\section{Design considerations}
Starting with the basic concepts of the Forest Flyer design, we discuss 
deficiencies and refinements to improve the key performance aspects of
projectile flatness (radial symmetry), projectile uniformity (axial symmetry),
and efficiency.
It is always desirable to know the condition of the projectile on impact as
accurately as possible.
For many shock physics experiments, it is desirable to induce a shock
which is as planar as possible over some region,
which is achieved by having the projectile as flat as possible, i.e. with
minimum deformation by the acceleration system.
High explosive launching systems must be designed with care, because the
pressures induced by detonation can easily exceed the flow stress of
materials, distorting the projectile if the flow of the detonation
products is not designed properly.
Similarly, it is usually desirable to generate a shock with a pressure
which is sustained at a constant value for a finite time.
This is achieved by preserving the projectile's mass density through its
thickness.
With the high launching pressures possible using explosives, it is
possible to induce large amplitude reverberating compression waves
and also reflected tensile waves which may open voids, reducing
the mean density of the projectile.
In contrast, projectiles launched conventionally by gas and powder guns
are subjected to accelerating pressures well below the flow stress of most 
solids, so projectiles generally remain flat and at their initial mass density.

\subsection{Choice of taper angle}
If the angle of tapering of the case is chosen incorrectly,
it could reflect a strong shock or rarefaction wave which could introduce
radial variations in loading, distorting the projectile radially.
For instance, a reflected shock could drive the outside of the projectile
harder than the region near the axis, resulting in a dish-shaped projectile
on impact.

The choice of the tapering angle depends on the explosive and the case
materials.
Ideally, the case would introduce no radial perturbation to the load
on the projectile.
This is generally not possible to arbitrarily high accuracy, 
as it would require that no wave be reflected
from the detonation wave or the subsequent release history.
Perturbations are likely to be greatest at the highest pressures,
so the most important requirement for a radially uniform drive is 
to ensure that no wave is reflected by the peak of the detonation wave.

The interaction of detonation waves with inert materials adjacent to the
explosive has been investigated in the context of detonation shock dynamics
\cite{Collyer98}.
The effect of the case depends qualitatively on whether it has a lower or
higher shock impedance than the explosive.

\subsubsection{Low-impedance case}
If the adjacent material has a lower shock impedance than the explosive,
it always reflects a rarefaction wave.
The effect of the case on the detonation wave can be
eliminated by choosing a taper angle such that the phase speed of the
interface is faster than the speed at which the rarefaction can propagate
across the detonation -- this is referred to as the `causal angle' $\phi_c$.

An observer sitting on a detonation wave sees unshocked material coming in at a speed $s$ and shocked material going out at $s-u$.
Signals emitted by a passing particle at a speed $c$ relative to the
particle move at a lower speed $c_w$ along the shock wave:
\begin{eqnarray}
c_w = \sqrt{c^2-(D-u_p)^2},
\end{eqnarray}
where $D$ is the speed of the detonation wave and $u_p$ the particle speed
at the shock front.
If the angle between the detonation wave and the edge of the explosive is $\phi$,
then the edge sweeps across the detonation wave at a speed
\begin{eqnarray}
c_s = \frac D{\tan\phi}.
\end{eqnarray}
The causal angle of incidence $\phi_c$ is the value when boundary effects
propagate just fast enough to influence other parts of the wave, i.e.
when $c_s = c_w$, so
\begin{eqnarray}
\tan\phi_c = \dfrac D{\sqrt{c^2-(D-u_p)^2}}.
\end{eqnarray}
A perfectly sonic shock would have $\phi_c=90^\circ$, requiring a taper
angle of zero.

\begin{figure*}
\begin{center}
\includegraphics[scale=0.7]{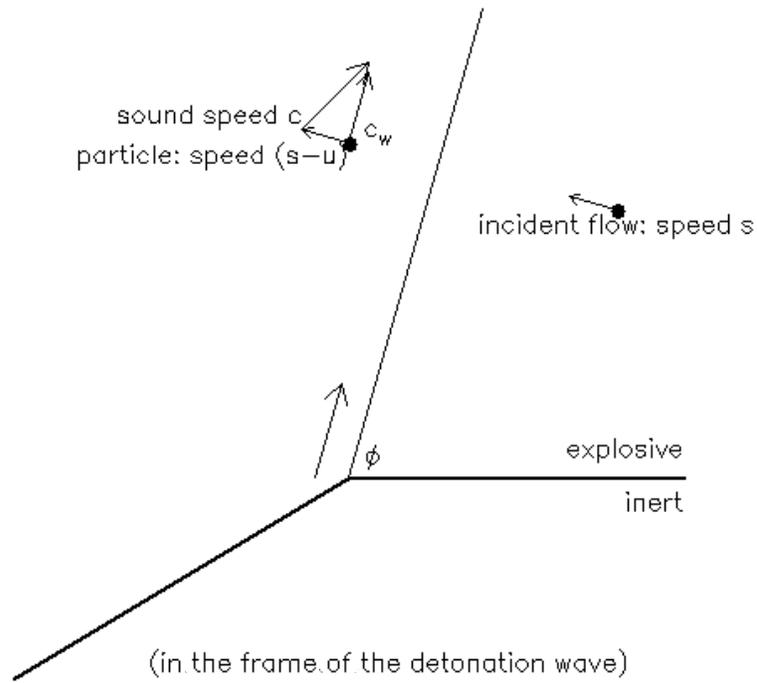}
\end{center}
\caption{Signals contributing to the causal angle $\phi_c$.}
\label{fig:caussig}
\end{figure*}

\subsubsection{High-impedance case}
If the adjacent material has a higher shock impedance than the explosive, 
it reflects a rarefaction wave for some range of angles
including the case of zero taper angle, and a shock at higher angles.
\footnote{%
For zero taper angle -- the detonation running parallel to the interface --
it can be seen intuitively that a rarefaction is reflected by
considering that, for one dimensional flow, 
the particle speed behind the detonation is parallel with the boundary,
whereas in the normal case where the shock speed in the case is lower
than the detonation speed, the particle speed in the case must have a radial
component, so a rarefaction wave must be reflected from the case in order
to impart a radial component to the reaction products and hence 
preserve continuity of radial velocity at the interface.}
The effect of the case can be eliminated by choosing the taper angle
such that no wave is reflected by the detonation -- this is referred to below
as the `asymptotic angle' $\phi_a$ as this is the angle at which a steady
detonation will make with the case beside a long charge.

Denoting quantities in the explosive and inert by a subscript `X' and `I'
respectively, and states before and behind the shock with `0' and `1', then
the following relations can be used to determine $\phi_a$:
\begin{itemize}
\item Continuity of the shock intersections with the boundary
(the `fundamental law of shock wave refraction' \cite{Henderson89}):
\begin{eqnarray}
\frac{s_X}{\sin\phi_X} = \frac{s_I}{\sin\phi_I},
\end{eqnarray}
where the $s$ are the speeds of the shock in each material
($s_X\equiv D$).
\item Flow deflection through a shock moving obliquely to a frame of
reference:
\begin{eqnarray}
\tan(\phi - \chi) = \frac{v_1}{v_0}\tan\phi,
\end{eqnarray}
derived from mass conservation (the ratio of particle speeds relative to the
shock normal is equal to the ratio of specific volumes $v_i$) and the
requirement that the shock does not alter the flow rate parallel to it.
This equation applies to both the explosive and the inert. They have their own
specific volumes and angles of incidence $\phi$, but the same angle of
deflection $\chi$.
\end{itemize}
The principal shock Hugoniot of both materials and the relations 
above are sufficient to determine $\phi_a$.
A perfectly rigid case would have $\phi_a=90^\circ$, requiring a taper
angle of zero.

\begin{figure*}
\begin{center}
\includegraphics[scale=0.7]{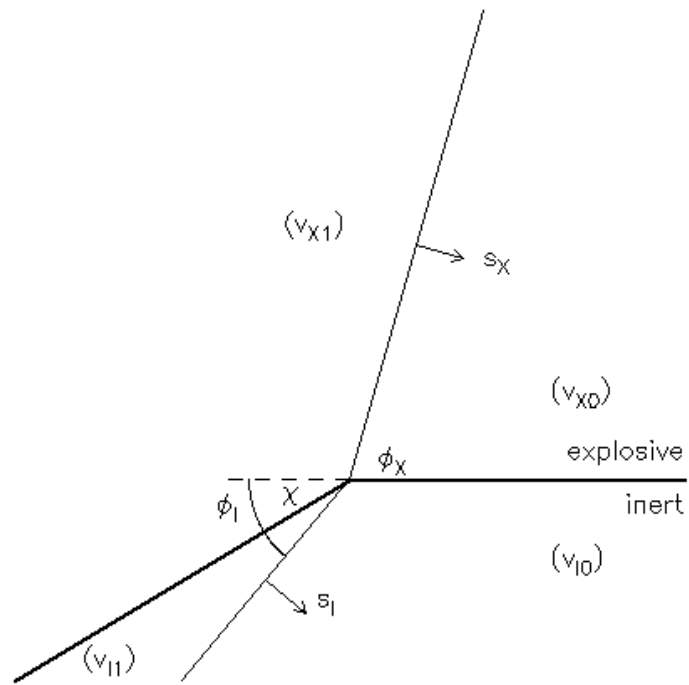}
\end{center}
\caption{States in the computation of the asymptotic angle $\phi_a$.}
\label{fig:asymstat}
\end{figure*}

\subsubsection{Application to explosively-launched projectiles}
Previous Forest Flyer variants have used a range of case materials
including ceramics, polymethyl methacrylate (PMMA), and Al alloy.
PMMA has a much lower shock impedance than the PBX-9501 charge,
and Al alloy has a higher impedance.
The taper angle used was 45$^\circ$.

In calculating the asymptotic angle, the choice of the
shocked state in the explosive is ambiguous.
Taking the laminar Zel'dovich-von Neumann-Doering (ZND) model of the detonation
process, detonation commences with an unreactive shock --
the von Neumann (vN) spike -- followed by reaction
as the explosive expands down the Rayleigh line to the fully-reacted
Chapman-Jouguet (CJ) state \cite{Fickett79}.
Physically, the effect of the case depends on the reflection from the
sequence of states in the axially-propagating detonation.
The CJ state is sonic with respect to the flow behind, so it is not
relevant for the calculation of the causal angle: the vN state must be used.
The vN state is not necessarily an accurate representation of the initial
shock in heterogeneous explosives such as PBX-9501, as the detonation wave is
not perfectly laminar.
It is also extremely difficult to measure the unreacted Hugoniot for
explosive materials to pressures approaching the CJ state, let alone higher
pressures, because chemical reactions take place promptly.
Previous studies \cite{Collyer98} have shown variations of a few degrees in
$\phi_a$, depending which state was taken,
with the CJ state agreeing more closely with experimental measurements.
Here, the CJ state was used.

The properties of PBX-9501 were represented by an equation of state (EOS) 
of the Jones-Wilkins-Lee type \cite{Fickett79} for the unreacted explosive
and the reaction products, with parameters
derived from cylinder expansion tests \cite{Souers93,McCluskey05}.
No unreacted parameters were found for PBX-9501, so parameters for the
similar HMX-based explosive PBX-9404 were used instead.
The vN state had mass density $\rho=2.56$\,g/cm$^3$, pressure $p=39.6$\,GPa,
particle speed $u_p=2.45$\,km/s, sound speed $c=12.17$\,km/s, 
and shock speed $D=8.776$\,km/s.
The CJ state had mass density $\rho=2.49$\,g/cm$^3$, pressure $p=37.0$\,GPa,
particle speed $u_p=2.28$\,km/s, sound speed $c=6.51$\,km/s, 
and detonation speed $D=8.776$\,km/s.
The causal angle $\phi_c$ is a property of the explosive, irrespective of
the case material so longer as it has a lower shock impedance.
For the EOS used, $\phi_c\simeq 40.2^\circ$.
The properties of Al alloy were represented by EOS of the 
Gr\"uneisen type using the principal Hugoniot as the reference curve,
with parameters derived from measurements of shock and particle speeds
from projectile impact experiments \cite{Steinberg96}.
The asymptotic angle $\phi_a$ was calculated as above for Al alloy,
and also for steel as a plausible alternative material providing greater
confinement of the reaction products
The asymptotic angle was 68.1$^\circ$ and 78.7$^\circ$ 
for Al and steel respectively, implying an ideal taper angle of
21.9$^\circ$ and 11.3$^\circ$ respectively.
Choosing these precise angles, the projectile may not be as flat as
possible because of other wave interactions and friction with the case
as the projectile accelerates, but these predictions should be accurate
to within a few degrees.

The predicted causal angle is just less than 45$^\circ$, 
so the Forest Flyer variant
with a 45$^\circ$ taper and a low impedance case should be
effective in reducing edge effects, and should hence give a reasonably 
flat projectile.
The predicted asymptotic angle for an Al alloy case is significantly greater than
45$^\circ$, so the Forest Flyer design with a 45$^\circ$ taper and 
an Al alloy case is potentially susceptible to problems from the reflected shock.

Continuum dynamics simulations were performed to evaluate the effect
of two dimensional reflections from the tapered case on the flatness of the
projectile.
The simulations were based on the design used for the neutron resonance
spectroscopy measurements \cite{Yuan05}:
a charge 37\,mm long and 89\,mm in diameter with a 45$^\circ$ taper to
63.5\,mm diameter,
a case 6.4\,mm thick,
a 7\,mm gap, and a projectile 6\,mm thick by 63.5\,mm diameter.
The simulations were Eulerian, using operator splitting between a 
second order Lagrangian phase of the predictor-corrector type and a third
order remapping using third order van Leer flux limiters and Youngs
interface reconstruction \cite{Benson92}.
Artificial viscosity of the von Neumann and Wilkins type was used to 
stabilize shock waves.
Explosive detonation was treated by constant speed programmed burn using
the CJ speed appropriate for the PBX-9501 EOS.
The materials used in the Forest Flyer are common, and their EOS and strength
properties have been investigated and reliable models developed
\cite{Steinberg96}.
The numerical methods used for the continuum dynamics simulations have been
used widely for explosive loading systems.
Hence the simulations should be broadly accurate.

Explosive initiation was simulated as a simultaneous plane over the
cylindrical face of the charge.
Convergence with respect to spatial resolution was tested, and found to be
around 0.1\,km/s in projectile velocity
for a resolution of $0.25 \times 0.5$\,mm.
The effect of material strength was evaluated by performing simulations
with an elastic-perfectly plastic strength model.
The speed and shape of the projectile were not significantly different.
The effects of tensile damage and spall were included through a
minimum pressure model \cite{Bushman93}.
With this model, unless the material is subjected large tensions for long
periods, it will recompress itself and `heal.'
No spall model could be found which was likely to be valid in the loading
history applied to the projectile.
The effect of tensile stress is discussed further below.

\begin{figure*}
\begin{center}
\includegraphics[scale=1.0]{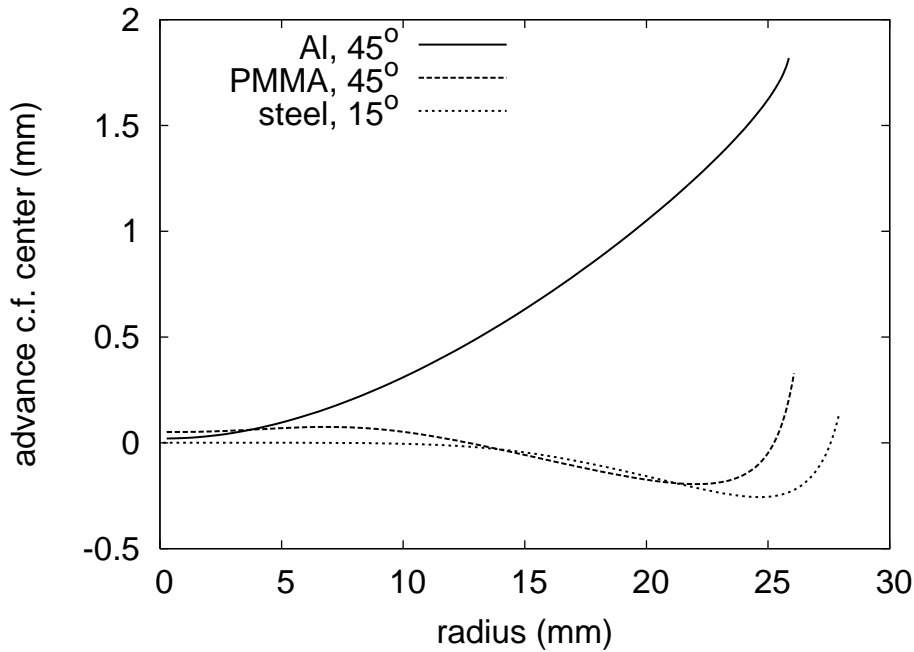}
\end{center}
\caption{Comparison of projectile flatness predicted from continuum dynamics
   simulations with different case materials and taper angles.
   45$^\circ$ tapers used the original Forest Flyer design with a 
   cylindrical section in the explosive charge;
   the other tapers were constant over the full length of the charge.
   The Al case with 20$^\circ$ taper gave much the same result as the steel.}
\label{fig:shape}
\end{figure*}

The shape and speed of the projectile were evaluated at an acceleration
distance of 20\,mm, where the target was placed in typical impact
experiments.
The projectile was accelerating and reverberating, so there was some
ambiguity in the predicted speed, represented as an uncertainty.
With a PMMA case, the projectile was calculated to be flat to within 
0.15\,mm over 31\,mm diameter.
With an Al alloy case, the projectile was predicted to travel slightly
faster, 
but to be significantly advanced at larger radii.
The speed with an Al alloy case matched the speed of $3.6\pm 0.1$\,km/s 
in the experiments
(Figs~\ref{fig:shape} and \ref{fig:uhistcmp}, and Table~\ref{tab:perf}.)

There is no compelling hydrodynamic reason to have a cylindrical section 
in the charge.
Simulations were performed of design variants with a constant taper over the
full length of the charge.
These simulations used the predicted asymptotic angles for Al alloy and
steel, adjusted slightly to improve the flatness of the projectile.
In both cases, the projectiles were calculated to be flat to better than
0.05\,mm over more than 30\,mm diameter.
With the Al alloy case, the modified design gave the same projectile speed
as the original design, for less explosive.
With the steel case, the explosive mass was significantly less again, 
for essentially the same projectile speed.
(Figs~\ref{fig:shape} and \ref{fig:uhistcmp}, and Table~\ref{tab:perf}.)

\begin{table}
\caption{Projectile acceleration performance for different cases and tapers.}
\label{tab:perf}
\begin{center}
\begin{tabular}{|l|c|l|r|r|}\hline
case & taper & initiation & charge mass & speed \\
 & (deg) & & (kg) & (km/s) \\ \hline
PMMA & 45 & multipoint & 0.387 & $3.5\pm 0.1$ \\
Al alloy & 45 & multipoint & 0.387 & $3.6\pm 0.1$ \\
Al alloy & 20 & multipoint & 0.308 & $3.5\pm 0.1$ \\
steel & 15 & multipoint & 0.286 & $3.5\pm 0.1$ \\
PMMA & 45 & plane wave lens & 0.996 & $3.8\pm 0.1$ \\
Al alloy & 45 & plane wave lens & 0.996 & $4.1\pm 0.1$ \\
\hline\end{tabular}

For the plane wave lens initiated charges, 0.367\,kg of the explosive mass
is TNT rather than PBX-9501.
\end{center}
\end{table}

\subsection{Initiation method}
One family of Forest Flyer designs uses an explosive lens to initiate the
charge over a plane.
The lens, comprising two explosives with different detonation speeds
to induce a flat detonation wave from a point detonator,
is a potential source of two dimensional perturbations, because
the pressure history induced varies radially, even if the initial pressure
and arrival time are simultaneous.

Continuum dynamics simulations were used to assess the effect of using
a plane wave lens (PWL) on the projectile shape and speed.
The effect on the projectile loading history was also assessed;
this is discussed later.
The PWL was a Los Alamos type P-108, comprising PBX-9501 as the fast component
and trinitrotoluene (TNT) as the slow \cite{Olinger06}.
PWL simulations are an interesting problem for continuum dynamics,
as there is potential for the slow explosive and `acceptor' charge to
be subjected to unusual conditions including overdriving, 
which are relatively poorly understood and may not be treated accurately by
explosive models generally used.
PWL simulations are discussed in more detail elsewhere \cite{Swift_pwl_07}.
Programmed burn with a constant CJ detonation speed was used in each
component.
The CJ detonation pressure in TNT is around 19\,GPa, which should induce
full detonation in PBX-9501 within a run distance of around 1\,mm.
This is short in comparison with the length scales of interest for the
Forest Flyers, so programmed burn is likely to be a good approximation here.
Overdriving of the TNT charge by the PBX-9501 component in the PWL is
a potential concern.
This was assessed by looking for any unphysical spike in the pressure history
exhibited in the TNT.
No significant spike was seen, indicating that constant speed programmed burn 
was unlikely to cause numerical problems although it does not capture the
full range of dynamical effects exhibited by real detonation waves in TNT,
specifically changes of speed caused by overdriving and curvature.

Using programmed burn with nominal CJ detonation speeds and JWL products EOS
\cite{Souers93} the P-108 lens was predicted to give a detonation arrival
which was 0.6\,$\mu$s later at the edges than on axis.
This deviation from perfect planarity was larger than predicted for
smaller PWL designs \cite{Swift_pwl_07}.
However, the shape of the Forest Flyer projectile was not significantly
different from the plane-initiated variant without the PWL,
although the projectile speed was predicted to be appreciably faster
with the larger mass of explosive (Table~\ref{tab:perf}),
consistent with the measured speed of $4.3\pm 0.3$\,km/s 
with a similar lens \cite{Buttler06}.

\subsection{Axial dimensions}
The family of explosive launchers considered here can be characterized 
as a radial configuration chosen to make the acceleration of the
projectile uniform in the radial direction,
and a set of component thicknesses in the axial direction.
The axial thicknesses are the explosive charge, the expansion gap between the
explosive and the projectile, the projectile, and the acceleration gap
between the projectile and the target.
If time-dependent material properties can be neglected, which is often the
case to an accuracy of a few percent for explosive systems, then it is the
ratio of thicknesses that matters, and the complete system can be scaled.

\begin{figure*}
\begin{center}
\includegraphics[scale=1.0]{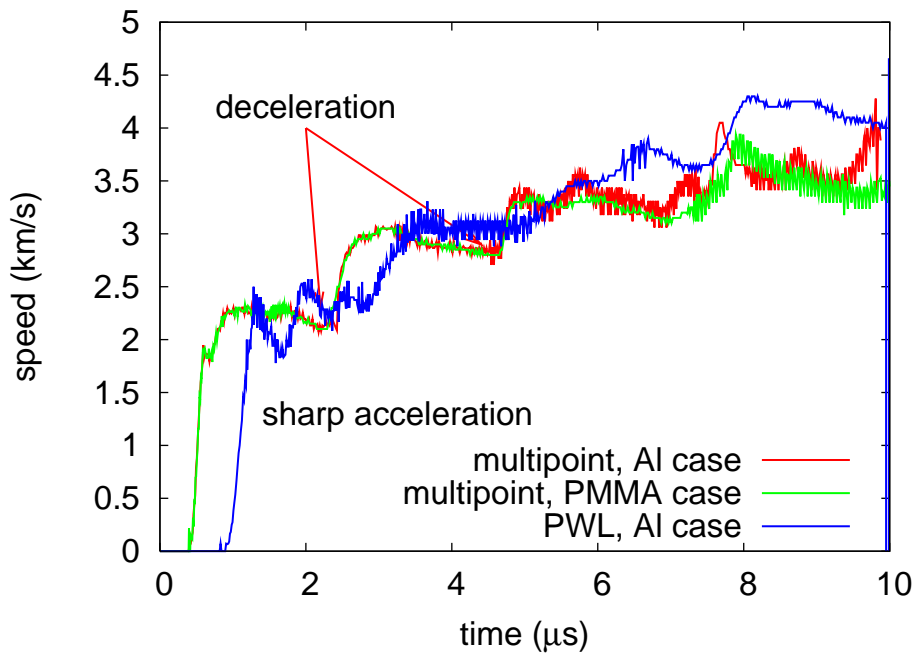}
\end{center}
\caption{Example free surface velocity histories from two dimensional
   continuum dynamics simulations of explosively launched projectiles,
   suggesting shock formation and tension.
   High frequency noise is from the use of numerical differentiation of the
   reconstructed interfaces to calculate the velocity in the Eulerian
   simulation.
   The time in each simulation was offset so that acceleration occurred at
   similar times in all cases.}
\label{fig:uhistcmp}
\end{figure*}

The two dimensional continuum dynamics simulations of the original
Forest Flyer designs and the designs with a modified taper but the
original axial thicknesses all exhibited free surface velocity histories
which were violent, suggesting the possibility of shock formation in the
projectile, and with decelerations, indicating tension
(Fig.~\ref{fig:uhistcmp}).

The axial design objective is to accelerate the projectile efficiently
to a constant speed without
inducing shock loading or tensile damage.
For a given thickness of explosive,
a smaller expansion gap means that the explosive products expand less before 
reaching the projectile, so a higher pressure is applied and the time from
the start of the drive to the peak pressure is shorter -- both resulting
in a higher pressurization rate $\dot p$ applied to the projectile.
The higher the pressurization rate, the shorter the distance within the 
projectile needed for the compression wave to steepen into a shock wave.
The peak tensile stress is induced when the accelerating pressure is reduced,
and the resulting rarefaction wave reflects from the front surface of
the projectile.
The faster the pressure is reduced, the stronger the tensile stress.
For a given projectile thickness, the expansion rate in the rarefaction 
can be reduced by increasing the thickness of the explosive or the 
expansion gap.
It is possible to express these sensitivities algebraically, but
this is of limited value as the material properties and wave
interactions are fairly complicated.
Instead, one dimensional, axial, continuum dynamics simulations were made of
the acceleration process and the loading history in the projectile.
One dimensional simulations allowed a finer spatial resolution,
and meant that a Lagrangian representation of the components could be
used, which is much more convenient when following the load and response of
individual elements of the material as opposed to particular locations in
space.
Again, a second order accurate time integration scheme of the 
predictor-corrector type was used, with artificial viscosity of the 
von Neumann and Wilkins type to stabilize shock waves \cite{Benson92}.
The spatial resolution used was 0.2\,mm in the explosive and 0.1\,mm in the
projectile.

\begin{figure*}
\begin{center}
\includegraphics[scale=1.0]{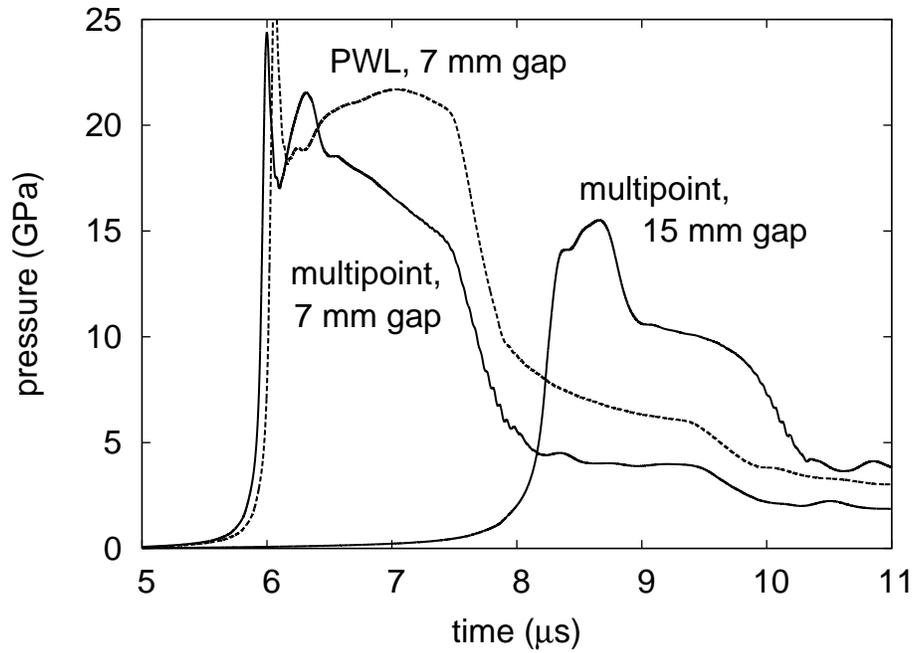}
\end{center}
\caption{Drive pressure histories predicted
   by one dimensional continuum dynamics simulations, for plane wave lens
   and multipoint initiation schemes, using the component thicknesses from
   the original Forest Flyer design and also for a thicker gap.}
\label{fig:drivecmp}
\end{figure*}

\begin{figure*}
\begin{center}
\includegraphics[scale=1.0]{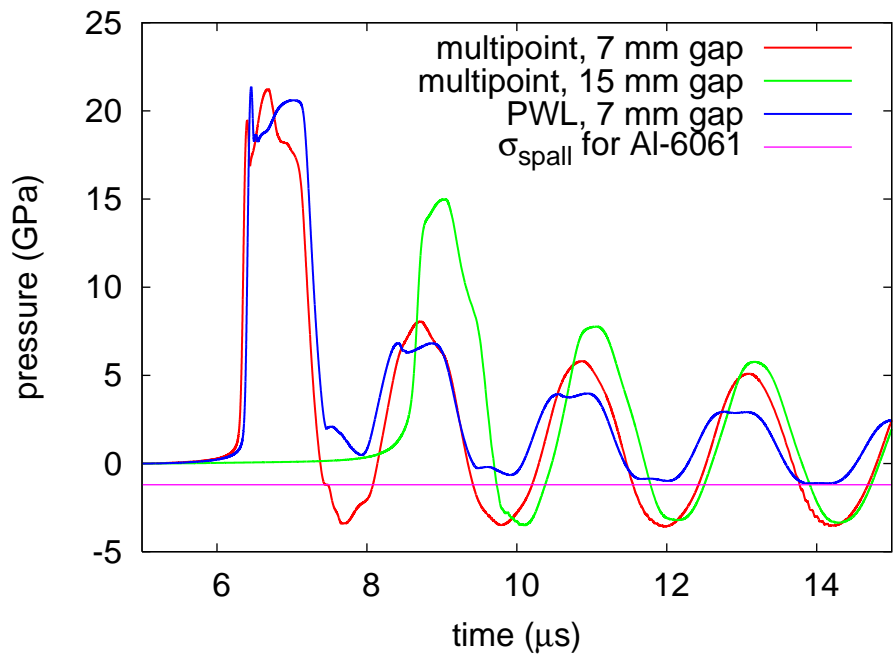}
\end{center}
\caption{Pressure histories predicted at the midplane of the projectile
   by one dimensional continuum dynamics simulations, for plane wave lens
   and multipoint initiation schemes, using the component thicknesses from
   the original Forest Flyer design, and also for a thicker gap.}
\label{fig:midcmp}
\end{figure*}

Simulations were performed representing multipoint and PWL initiation.
For PWL, an additional thickness of 67\,mm of TNT was used to
represent the PWL.
The one dimensional representative thickness should be less than the
actual thickness to account for radial expansion and the lower pressure
imparted by the TNT component.
%A reasonable thickness is the radius of the PWL.
The free surface velocity history was consistent with the two
dimensional simulations and experiments.
The drive pressure applied to the projectile rose to over 20\,GPa
in the original Forest Flyer designs, and as intended was shockless
(Fig.~\ref{fig:drivecmp}).
The loading history was investigated at the center of the projectile.
With PWL initiation, the simulations predicted that the tensile stress would
just reach the spall strength of Al alloy inferred from projectile impact
experiments \cite{Steinberg96};
with multipoint initiation, the faster decrease of pressure gave a significantly
higher tensile stress (Fig.~\ref{fig:midcmp}).
The damage and spall response of materials depends on the strain rate,
with generally higher strength being exhibited at higher strain rates.
The strain rates experienced by the projectile in these experiments was
considerably slower than in typical projectile impact experiments or in
loading by an explosive in direct contact, so it would be reasonable
to expect more damage than implied by the published spall model,
though it is difficult to predict the precise amount without a more
sophisticated understanding of dynamic damage than exists at present.
Thus, with the component thicknesses used, the Forest Flyer variants should be
considered as unsafe in respect of tensile damage to the projectile,
and it would be surprising if the multipoint design did not induce
a large amount of tensile damage.
The accelerating pressure from the expanding reaction products is sustained
for several cycles of compression and tension, so some recompaction is likely
to occur.

Given the pressurization rate $\dot p(t)$ applied to the projectile,
the distance required for the resulting compression wave to steepen into
a shock is the distance at which adjacent characteristics intersect.
If at some instant $t$ the interface has been compressed to a displacement $x$,
and the particle speed and longitudinal sound speed are $u$ and $c$
respectively, adjacent characteristics $u+c$ intersect at a distance
\begin{equation}
x_{int}=x+\frac{c(c+u)}{d(c+u)/dt},
\end{equation}
where
\begin{equation}
x=\int u\,dt,
\end{equation}
$c$ and $u$ are calculated as a function of $p$ along the isentrope
of the projectile material,
and the derivative is calculated from $\dot p(t)$ applied along the
isentrope.
This analysis assumes no release from the front of the projectile,
so it is suitable for finding the thickest projectile that can be driven
shocklessly with a given loading history.
Analysis using the intersection of characteristics is more appropriate than
the continuum dynamics simulations because the use of artificial viscosity
to stabilize shock waves makes it difficult to distinguish them from
steep but shockless compression waves.

\begin{figure*}
\begin{center}
\includegraphics[scale=1.0]{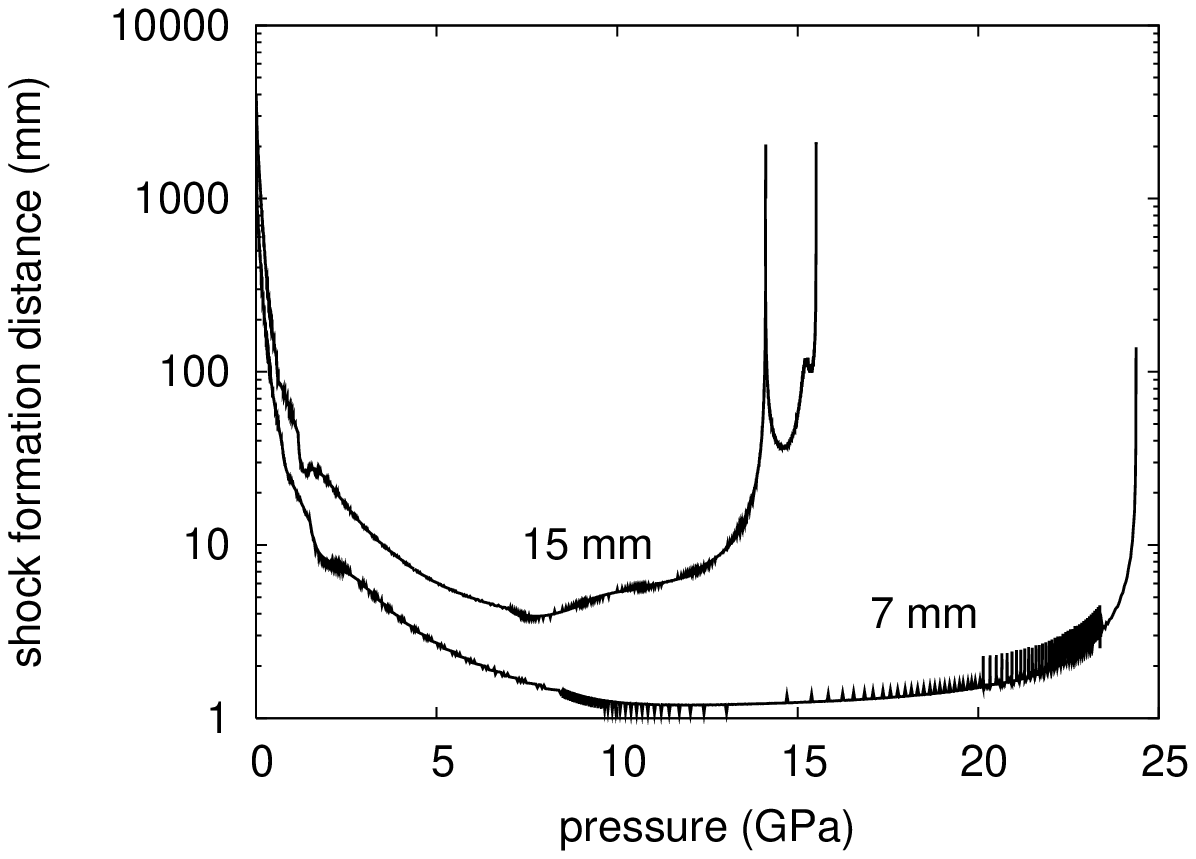}
\end{center}
\caption{Shock formation distances predicted as a function of drive
   pressure as predicted from
   one dimensional continuum dynamics simulations with
   multipoint initiation, using the component thicknesses from
   the original Forest Flyer design (7\,mm gap)
   and also for a thicker gap (15\,mm).
   The high frequency noise is from the use of a tabulated numerical calculation
   of the isentrope.}
\label{fig:shockform}
\end{figure*}

With the 7\,mm expansion gap, the load applied to the projectile rose to over
20\,GPa.
Considering the intersection of characteristics, the compression wave 
steepened into a shock for pressures from 10 to 20\,GPa in around 1.5\,mm,
so the projectile was substantially shock-loaded.
A single shock to 20\,GPa and release to zero pressure in Al 
produces a residual temperature of around 400\,K
taking plastic flow into account,
and a 1\%\ reduction in mass density.
The shock formation distance was slightly shorter at intermediate pressures
using a plane wave lens instead of multipoint initiation.
With a 15\,mm expansion gap, the maximum load was around 15\,GPa and the
pressurization rate was significantly lower, with the shortest shock
formation distance around 4\,mm for pressures around 7.5\,GPa and 
a shock forming between 6 and 10\,GPa for a projectile 6\,mm thick;
this would lead to much less residual heating.
(Fig.~\ref{fig:shockform}.)

\begin{figure*}
\begin{center}
\includegraphics[scale=1.0]{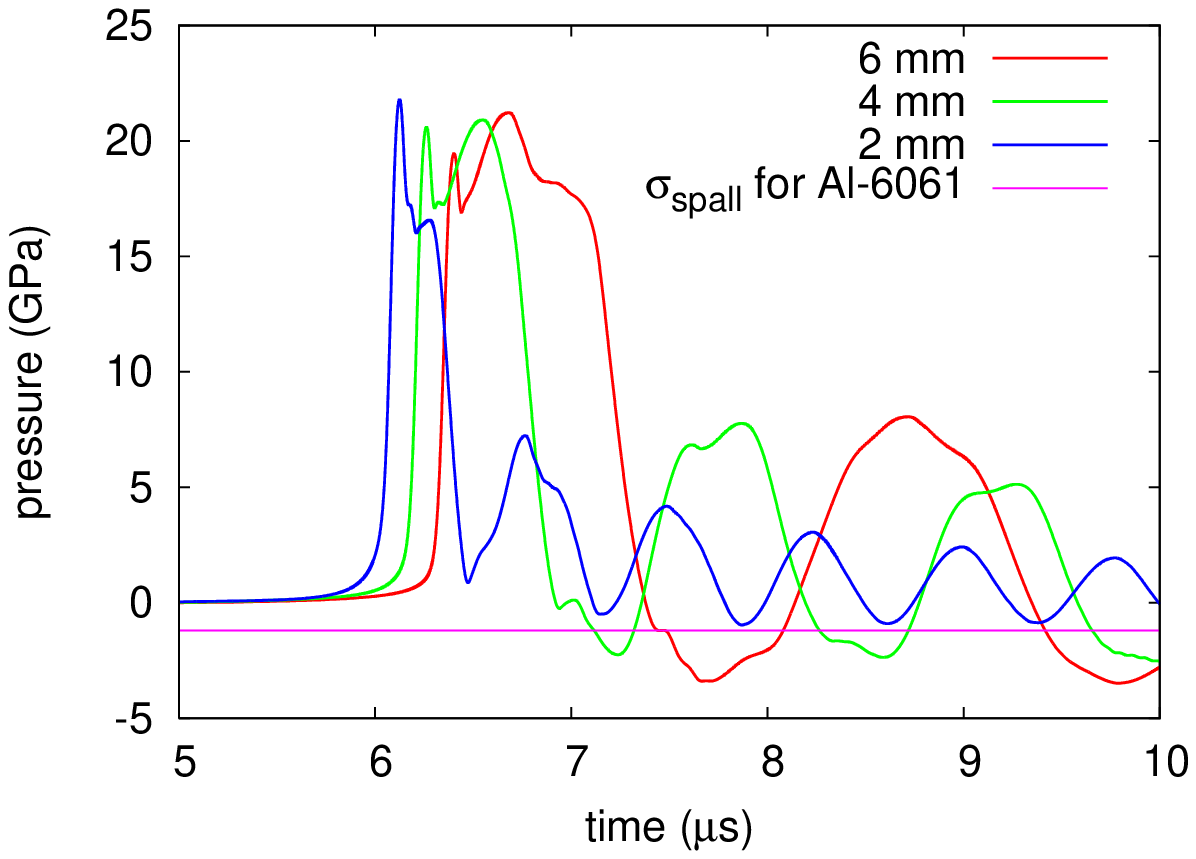}
\end{center}
\caption{Sensitivity of loading history at midplane of projectile
   to projectile thickness.
   Explosive 37\.mm thick, 7\,mm gap, multipoint initiation.}
\label{fig:thickcmp}
\end{figure*}

Simulations were used to quantify the effect of changing the thickness of
the expansion gap and the projectile, with respect to the explosive.
The magnitude of the tensile stress was relatively insensitive to the
gap thickness, though it could be reduced quite easily by reducing the
projectile thickness, also giving a higher projectile speed.
(Fig.~\ref{fig:thickcmp}.)

\begin{figure*}
\begin{center}
\includegraphics[scale=0.6]{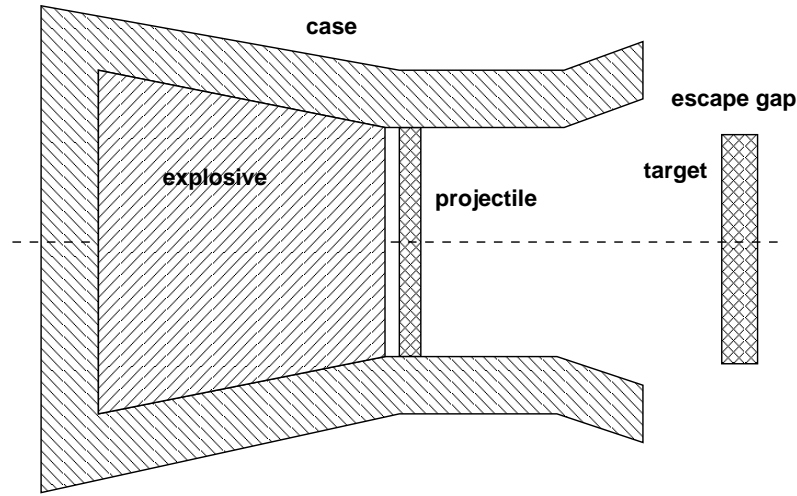}
\end{center}
\caption{Schematic of optimized explosive launching design with thick,
   constant taper case of high impedance material such as steel,
   and a radial escape gap to reduce drive gradients in the
   projectile before impact.}
\label{fig:escapeschem}
\end{figure*}

The acceleration rate increases rapidly at first to a maximum
when the peak pressure is applied, then falls steadily as the 
explosive products expand.
Lateral expansion, i.e. two dimensional effects, reduce the acceleration
at late times.
The greater the thickness and mass density of the case, the less
the energy used to accelerate it, and the greater the efficiency of
projectile launch.
If the projectile is accelerating, it has a gradient of pressure
and mass density through its thickness.
To induce a shock on impact which is as constant as possible in time,
the axial gradients should be as small as possible.
For a given thickness of explosive, expansion gap, and projectile,
the projectile speed is higher and closer to constant for a larger
acceleration gap.
However, the larger the gap the greater the effect of radial variations in
pressure which may deform the projectile.
It may be useful to include a radial escape gap for the reaction products 
just before the target, to reduce the axial gradients in the projectile
(Fig.~\ref{fig:escapeschem}).
If the acceleration gap is too small, the higher density reaction products
may scour the rear of the projectile as they expand laterally, deforming it.

\section{Experimental observations of flatness and damage}
A conclusion from the hydrodynamic analysis and simulations is that,
with an Al alloy case, the
original Forest Flyer design would be expected to give a dish-shaped
projectile, and this can be avoided by a more careful choice of case taper
angle.
The experiments performed to investigate the original designs measured
the projectile's flatness from its impact with a flat glass block,
with Ar-filled grooves which produce a light flash when compressed
\cite{Forest98,Funk97}.
These experiments suggested that impact was more planar than predicted by the
simulations, though they did indicate that the edges of the projectile arrived 
earlier than the axis.

The Ar-flash technique indicates when a substantial amount of material has
arrived, but it does not guarantee that the projectile is solid.
Since the simulations also predicted that the projectile was subjected to
a strong axial tension, a plausible explanation is that the projectile
was damaged by the opening of internal voids,
%and possibly even a complete spall
%plane (Fig.~\ref{fig:damageschem}), 
for multipoint initiation with an Al alloy case.

%\begin{figure*}
%\begin{center}
%\includegraphics[scale=0.6]{damageschem.eps}
%\end{center}
%\caption{Schematic of internal tensile damage to the projectile,
%   proposed to reconcile the difference between continuum dynamics
%   simulations with no or a simple spall model, and experimental
%   measurements indicating a flatter face to the projectile,
%   for Forest Flyer systems using an Al alloy case and multipoint initiation.}
%\label{fig:damageschem}
%\end{figure*}

Explosive launching of projectiles does not present particular problems
for continuum dynamics simulations.
The states induced in the materials are not notably different from states
induced in other explosively-driven systems.
The materials used in the Forest Flyer designs are not unusual or
poorly-understood.
One would expect continuum dynamics simulations to be accurate, certainly
for the broad deformation and motion of the components.
Indeed, the simulations reproduced the measured projectile speeds to within
their uncertainty.
Damage and spall depend more sensitively on the precise composition and
manufacturing history of a component, and the loading history applied.
A discrepancy related to tensile damage is quite possible, and could well
explain why the arrival-time measurements suggested that the projectile
was flat.
It would be valuable to perform projectile acceleration and impact trials
with a more detailed diagnostic such as imaging Doppler velocimetry or 
radiography,
to measure or infer a density distribution through the projectile.

\begin{figure*}
\begin{center}
\includegraphics[scale=0.6]{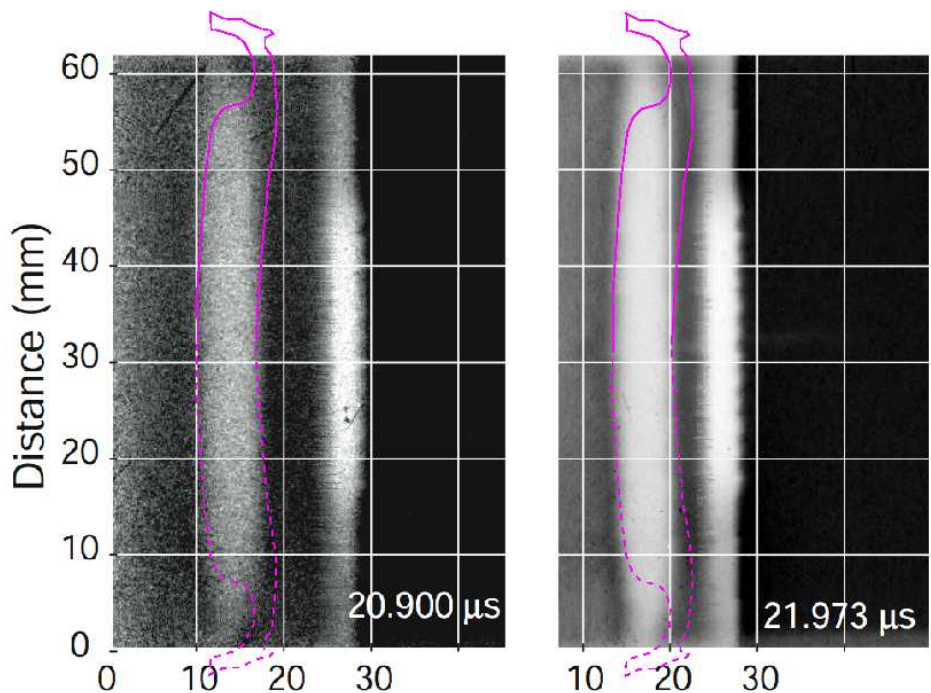}
\end{center}
\caption{Comparison between proton radiographs showing Forest Flyer shape
   just before impact and projectile shape from continuum dynamics simulations.
   The calculated curvature of the rear surface matches well over the central
   40\,mm.  The thin fillet at the outside is visible further back in the
   radiograph.  The leading edge at the outside of the projectile appears
   further advanced in the simulations.}
\label{fig:pradcmp}
\end{figure*}

Fortuitously, proton radiographs were obtained of a Forest Flyer projectile
just before impact in experiments to study surface ejecta from Sn 
\cite{Buttler06}.
In these experiments, the charge was initiated by a PWL, so the tensile
stress in the projectile was predicted to be less than with multipoint 
initiation.
The case was Al alloy, so the simulations predicted a reflected shock
from the tapered section and dishing of the projectile.
The radiographs were not optimized for reconstruction of the mass density in
the projectile, but they clearly showed curvature of the rear surface of
the projectile.
Simulations of the operation of this variant of the Forest Flyer were
in good agreement with the measured shape of the projectile,
except at the outside edge (Fig.~\ref{fig:pradcmp}),
indicating that two dimensional flow was predicted accurately and hence
that the predicted deformation was likely to be accurate.
It would be preferable to compare an unfolded density map from the
radiograph or a simulated radiograph from the simulation, but this is not
worthwhile without optimizing the radiography for the range of areal mass
in the projectile.
The front of the projectile is less likely to be accurate as this is the
region most affected by tensile damage.

\begin{figure*}
\begin{center}
\includegraphics[scale=1.0]{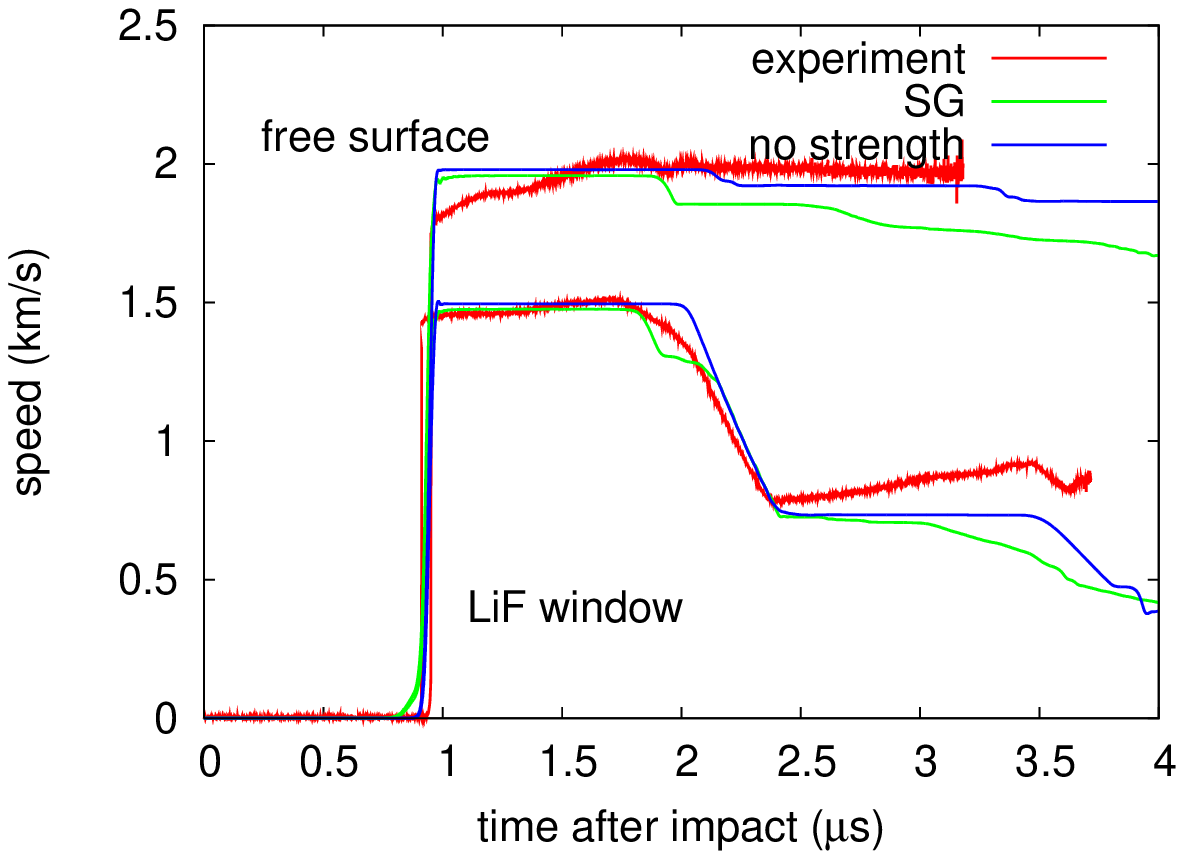}
\end{center}
\caption{Comparison between Doppler velocimetry records at the surface
   of a Mo target with continuum dynamics simulations
   with and without strength (SG: Steinberg-Guinan \cite{Steinberg96}),
   assuming a solid projectile.
   The low initial speed in the experimental records may
   indicate a region of low density near the front of the projectile,
   as may the smoother onset of elastic release from the peak velocity
   which shows up clearly in the LiF release record.}
\label{fig:visarcmp}
\end{figure*}

Doppler velocimetry measurements were made of the surface of a Mo target
impacted by a Forest Flyer driven Al alloy projectile, using multipoint
initiation \cite{Holtkamp06}.
In different experiments, measurements were made of the free surface
and of the surface in contact with a LiF window.
If the projectile had a constant mass density through its thickness,
the peak surface velocity induced by the shock should be constant.
The measured velocity histories showed acceleration during the peak,
which is a clear indication of density variations in the projectile;
these could be caused partly by the continued acceleration of the 
projectile in impact, but are likely to be caused mostly by damage-related
porosity in the projectile.
(Fig.~\ref{fig:visarcmp}.)

Another variant of the Forest Flyer design has been used to study
interface friction \cite{Marr-friction}.
The case was Al alloy,
the charge was initiated with a PWL, and the projectile was 15\,mm thick.
The axial analysis above predicted tensile stresses exceeding the spall
strength of Al-6061.
Two dimensional simulations predicted less curvature than the 6\,mm thick
projectiles.
A proton radiograph was obtained just after impact with the target assembly,
before the shock wave had reached the rear part of the projectile.
The radiograph clearly showed cracking of the projectile.
This observation correlates well with the simulations: damage
is likely to occur to the projectile when the tensile stress exceeds the
published spall strength, and quite possibly not for smaller tensile stresses.

\section{Discussion}
This study has shown that there are some subtleties in the hydrodynamic design
of explosive launching systems for projectiles.
Because the pressures induced by detonation waves and the expanding fluid
of reaction products can easily exceed the compressive and tensile strength 
of solids, care is needed to ensure that projectiles remain flat and solid.

There are compelling indications that the projectiles used in the 
neutron resonance temperature experiments on Mo were somewhat curved and 
damaged.
This deformation is of concern in experiments where the region to be studied
in the target is close to the impact face and when the state of interest is the
impact-induced shock, as in the case of the Mo shock temperature experiments.
The deformation is less important further through the target and when studying
states later in time, as in release temperature and surface ejecta experiments.
The projectiles can be made flatter by adjusting the shape of the case
around the explosive. Case improvements can also improve the
efficiency of the drive, reducing the amount of explosive needed or
increasing the maximum possible projectile speed.
There seems no reason not to use the case modifications for future
experiments based on the Forest Flyer design.

All Forest Flyer variants used so far are likely to subject the projectile
to tensile stresses around or in excess of its spall strength.
The expansion time scale of the detonation products,
and hence the time over which the projectile accelerates,
is longer than the reverberation time in the projectile, so 
material ejected from the surface of the projectile 
by the initial strong tension is likely to be recollected, and
material affected by tension-induced porosity may be recompressed,
but the projectile should not be assumed to comprise pristine material at
its initial density.
If the explosive charge is initiated with a plane wave lens, the explosive
in the lens reduces the expansion rate of the reaction products,
reducing the tensile stress compared with multipoint initiation,
but the tensile stress is still large enough to potentially cause damage.
The peak tensile stress can be controlled by adjusting the relative
thickness of the explosive, expansion gap, and projectile.

The use of arrival time images as a measure of projectile flatness is
limited, as it does not indicate how solid the projectile is
behind the leading material.
It is preferable to follow the acceleration history of the projectile
using Doppler velocimetry imaged along a line or even over an area
\cite{Tierney-2Dvisar}, complemented with lateral radiography if the
possibility of tensile damage is suspected.
The glass `witness block' technique could be used in conjunction with
Doppler velocimetry to measure the velocity history of the impact interface
instead of merely the arrival time;
this velocity history would provide a fairly direct indication of 
tension-induced porosity in the projectile.

The work reported here provides a deeper understanding of the physics
of projectile acceleration by explosives.
Using the Forest Flyer design as a basis, the radial and axial modifications
discussed, and the hydrodynamic details elucidated,
should lead to a family of projectiles giving loading conditions
closer to the one-dimensional ideal desirable for experiments in shock physics.

\section{Acknowledgments}
David Funk, Rob Hixson, and Ron Rabie provided information on the previous
Forest Flyer experiments.
We would also like to recognize the contributions of Langdon Bennett,
John Vorthman, Howard Stacy, and Dennis Shampine 
to the original Forest Flyer work.
Bard Olinger provided information on plane wave lenses.
Achim Seifter provided data from Forest Flyer experiments
with velocimetry and pyrometry diagnostics.
Chris Morris and Cynthia Schwartz, and the proton radiography team,
were instrumental in obtaining the radiographs of the Forest Flyer variant.
Partial funding was provided by the LANL Program Office for 
Laboratory-Directed Research and Development under an
Institutional Program Development project run by 
Kurt Schoenberg.
The work was performed under the auspices of
the U.S. Department of Energy under contracts W-7405-ENG-36
and DE-AC52-06NA25396.

%\clearpage

\end{document}